\documentclass[runningheads]{Springer/svmult}
\usepackage{makeidx}
\usepackage{graphicx}
\usepackage{Springer/subeqnar}
\usepackage{multicol}
\usepackage{Springer/physmubb}
\usepackage{amssymb}
\usepackage{amsmath}
\makeindex 
\newcommand{\greeksym}[1]{{\usefont{U}{psy}{m}{n}#1}}
\newcommand{\umu}{\mbox{\greeksym{m}}}

\newcommand{\Smo}{Smo\-lu\-chow\-ski}
\newcommand{\frag}{frag\-men\-ta\-tion}
\begin{document}

\title*{Kinetics of Fragmenting Freely Evolving Granular Gases}
\toctitle{Kinetics of Fragmenting 
\protect\newline Freely Evolving Granular Gases}
\titlerunning{Fragmenting Granular Gases}
\author{Ignacio Pagonabarraga, \\ 
Departament de F\'{\i}sica Fonamental, Universitat de Barcelona\\ 
Carrer Mart\'{\i} i Franqu\'es, 1,
08028-Barcelona, Spain\\
and \\
Emmanuel Trizac\\
Laboratoire de Physique Th{\'e}orique (UMR 8627 du CNRS), 
B{\^a}timent 210, Universit{\'e} de Paris-Sud,
91405 Orsay Cedex, France   
}

\authorrunning{I. Pagonabarraga and E. Trizac}
\maketitle
\abstract{
We investigate the effect of \frag\  on the homogeneous
free cooling of inelastic hard spheres, using Boltzmann kinetic theory
and Direct Monte Carlo simulations. We analyze in detail a model where
 dissipative
collisions may subsequently lead to a break-up of the grains.
With a given probability, two off-springs are then created 
from one of the two colliding partners, with conservation of mass,
momentum and kinetic energy. We observe a scaling regime characterized 
by a single collisional average, that quantifies the deviations 
from Gaussian behaviour for the joint size and velocity distribution
function. We also discuss the possibility of a catastrophe whereby 
the number of particles diverges in a finite time. This phenomenon appears
correlated to a ``shattering'' transition marked by a delta singularity at vanishingly small grains for the rescaled size distribution. 
}

%%%%%%%%%%%%%%%%%%%%%%%%%%%%%%%%%%%%%%%%%%%%%%%%%%%%%%%%%%%%%%%%%%%%%%%
\section{Introduction}

The phenomenon of fragmentation is central to many natural and technological
processes, from  geology to the pharmaceutical industry (where e.g. the properties
of an active principle depend on the size distribution of its constituents). 
In the realm of granular gases where the fluidization is achieved by a violent
excitation or shaking, the fragmentation mechanism is also relevant although 
experimental studies of these systems usually avoid such a complication.

A theoretical description of \frag\ is nevertheless a difficult task and traditional
approaches rely on rate (\Smo -like) equations \cite{redner}, that describe the 
time evolution of the size distribution after having integrated out the 
other degrees of freedom, such as the velocities of the particles. However, when 
the motion between the collision events is free, the dynamics of a reacting  system
in general (where, because of collisions, 
particles may break up, aggregate, annihilate \ldots) 
is determined by details of the collision (collisional correlations \cite{emm})
which often invalidate the rate equation approach, at least in its usual
version. In particular, for dynamics that do not conserve the number of particles,
the time evolution of the density crucially depends on the ratio of the 
kinetic energy dissipated in a ``typical'' collision, to the mean kinetic energy 
of the particles \cite{emm,pre}. It therefore appears that a more 
microscopic description
than the coarse grained rate equation route is necessary, at the expense
of a non negligible complication. 
When particles disappear upon colliding, it is however possible to prove
that the Boltzmann equation becomes exact at late times (in space dimensions
$D>1$ \cite{pre}). Although it does
not seem possible to show that the validity of Boltzmann's picture
(which is less restrictive than the \Smo\ approach)
pertains to a modification 
of the reaction rule, such an approach captures important collisional correlations
missed by the naive rate equation framework. In the following analysis, 
we will therefore use Boltzmann's kinetic theory 
(relying on  the molecular chaos assumption that the 2 body 
distribution function of colliding pairs may be factorized in terms 
of the single particle distribution \cite{Resibois}), to investigate 
the dynamics of a fragmenting granular gas.

%%%%%%%%%%%%%%%%%%%%%%%%%%%%%%%%%%%%%%%%%%%%%%%%%%%%%%%%%%%%%%%%%%%%%%%
\section{Model}
\label{sec:model}
We consider a gas of  grains that evolve freely in time, from an initial 
equilibrium distribution, i.e. we assume that up to the initial time, the 
system is  elastic, and hence, that its velocity distribution is Maxwellian.

In the fluidized regime we shall be interested in, 
grains interact as hard bodies; collisions are 
always binary, and instantaneous. In a $D$-dimensional system, at each collision a 
fraction $(1-\alpha^2)/(2 D)$ of the relative kinetic energy is lost, 
where $\alpha$, the inelasticity (or restitution) 
coefficient, measures the departure from  
elastic behaviour. After each collision, 
the stored kinetic energy in internal grain excitations may lead to a break-up 
of the grains. Such a \frag\ event is stochastic in nature. Previous 
studies suggest than in a number of materials, the probability that a grain breaks 
is related to the number of surface defects (or micro-cracks). The more 
kinetic energy is dissipated, the larger the chance that one of such defects 
enlarges, leading to the break-up of the grain. Hence, the fracture 
probability is related both to the energy lost and to the defect distribution
 (and the related surface stress distribution). There exist different models 
in the literature that account for such effects. A reasonable fragmentation 
probability model assumes that the stored energy and the defect distribution 
are uncorrelated, and that defects are uniformly distributed over a 
particle's surface; the breaking probability is thus 
proportional to the particle's 
surface \cite{poschel}. Once a grain breaks, we should also specify the number 
of out-coming particles, and how their sizes and velocities are related to 
the physical properties of the fragmenting grain.  Here the situation is more 
controversial, and it is highly dependent on whether the breaking process is 
induced by an energy input from a source which drives the system or whether it 
is due to the internal dynamics of the system; despite this, 
the Roslin-Rammler law (and a few variants) is considered to 
be a reasonable description for the off-spring distribution\cite{redner,poschel}.

Rather than attempting a detailed description of the \frag\ mechanism,  in 
this paper we will assume the simplest  break-up process. The idea is to 
minimize the amount of input at the collision level and to analyze  which 
is the behavior of the system under the simplest possible mechanical rules. We hope 
to be able to carry out a detailed analysis that will be helpful before
investigating more refined approaches. 

In our model, after the collision has taken place, with a probability $p$ 
we break one of the two colliding grains. The grain fragments in 2 
off-springs which keep the (spherical) shape of the parent grain. The \frag\ 
process conserves mass,  momentum and kinetic energy. The mass is distributed randomly
between the two out-coming grains with a uniform probability density. In 
this case, the conservation of momentum and kinetic energy imply 
that both off-springs have the same velocity, equal to that of the 
fragmenting grain. If we would  assume that the two colliding 
grains may break up due to the collision, there would exist more freedom in the 
velocity redistribution of the off-springs, even assuming coplanarity of the 
out-coming grains' trajectories. We will consider that the mass density of the grains
is conserved, so that the relation between mass $m$ and radius $\sigma$ 
is taken as $m = \sigma^D$. 

We will further concentrate on the homogeneous freely evolving state of the 
fluid under consideration. Such a
situation has received much attention in the absence of fragmentation, and the 
theoretical predictions have been contrasted against molecular dynamics 
simulations. The homogeneous state is known to be unstable against low 
wave length fluctuations of the vorticity, and at late times, hydrodynamic 
flows set up\cite{goldhirsch}. As a first step, we will concentrate in the regime prior to the 
appearance of the hydrodynamic instability. In fact, having a hydrodynamics 
origin, we might also expect an analogous instability in the present model.

In this homogeneous state, the  Boltzmann equation, which determines, at a 
mean-field level, the time evolution of the system through the one-particle 
distribution function $f({\vec v},\sigma,t)$, in a $D$-dimensional space  
has the form
\begin{eqnarray}
&\frac{\displaystyle\partial f({\vec v},\sigma,t)}
{\displaystyle\partial t} &= \frac{1}{2} \int du g(u)\int
 d{\vec v}_1 d{\vec v}_2d\sigma_1d\sigma_2 d\widehat{\varepsilon}\;\;
\theta({\vec v}_{12}\cdot\widehat{\varepsilon})\left|{\vec v}_{12}\cdot\widehat{\varepsilon}\right|
 \sigma_{12}^{D-1}  \left\{\right.\nonumber\\
&&f({\vec v}_1,\sigma_1,t)f({\vec v}_2,\sigma_2,t)\left[
-\delta({\vec v}-{\vec v}_1)\delta(\sigma-\sigma_1)-\delta({\vec v}-{\vec v}_2)
\delta(\sigma-\sigma_2)\right.\nonumber\\
&+&(1-p)\delta({\vec v}-{\vec v}_1^{*})\delta(\sigma-\sigma_1)+(1-p)
\delta({\vec v}-{\vec v}_2^{*})\delta(\sigma-\sigma_2)\nonumber\\
&+& p\delta({\vec v}-{\vec v}_1^{*})\left[\delta(\sigma-\sigma_1\cdot
 u^{1/D})+\delta(\sigma-\sigma_1\cdot
 (1-u)^{1/D})\right]\nonumber\\
&+&\left.\left. p\delta({\vec v}-{\vec v}_2^{*})\left[\delta(\sigma-\sigma_2\cdot
 u^{1/D})+\delta(\sigma-\sigma_2\cdot
 (1-u)^{1/D})\right]
\right]\right\}
\label{eq:boltzm}
\end{eqnarray}
In the previous expression, ${\vec v}$ stands for the velocity, $\sigma$ for
 the particles' radius, while $\widehat{\varepsilon}$ is a unit vector joining the
centers of the colliding grains. The Heaviside 
function, $\theta$, ensures the appropriate kinematic constraint, 
 that only approaching grains will collide, while the term
 $\left|{\vec v}_{12}\cdot\widehat{\varepsilon}\right| \sigma_{12}^{D-1}$ is the collision 
cross-section; the subscript $12$ refers to the difference between the values 
of the colliding pair, e.g. ${\vec v}_{12}={\vec v}_1-{\vec v}_2$. Moreover,
 $g(u)$ is the probability density that one of the off-springs, in case of 
fragmentation, has a mass $u$ times the parent's mass. Due to mass
 conservation, $g(u)=g(1-u)$. In the model analyzed below, 
 we consider the simplest case where this distribution is uniform [$g(u)=1$
and $0 \leq u \leq 1$].

For clarity's sake, we have expressed the general evolution equation in 
terms of the elementary processes affecting the two colliding grains. 
Obviously, a simplified expression can be derived using the symmetric role 
that the two grains play during the collision.

The first two terms in the r.h.s. of (\ref{eq:boltzm}) account for 
the disappearance of particles of a given velocity due to the collision 
process itself. For the same reason, new particles of a given species and 
velocity are produced at the collision. These terms involve the precollision 
velocities,  ${\vec v}^{*}$, because we require that the out-coming velocity of the 
collision corresponds to the velocity ${\vec v}$. It is through these, 
precollisonal velocities, that the inelastic character of the collision enters.
 Indeed,  the precollision velocities, in terms of the post-collisional ones 
(which are the relevant velocities in the previous kinetic equation in order 
to describe the kinetic processes), are given by 
\begin{equation}
{\vec v}_{1,2}^{*}={\vec v}_{1,2}\mp\frac{m_2}{m_1+m_2}\left(1+\alpha^{-1}\right)
\left[\,\widehat{\varepsilon}\cdot{\vec v}_{12}\,\right]\widehat{\varepsilon}
\label{eq:collision}
\end{equation}
The two terms in (\ref{eq:boltzm}) containing the prefactor $1-p$ 
correspond to the collisions in which, with probability $1-p$, particles do 
not break. The last four terms account for the fragmentation of grain 1 with mass 
$m_1 (= \sigma_1^D)$ (resp. grain 2 with mass $m_2$), leading to the
 creation of two off springs
with masses $u m_1$ and $(1-u) m_1$ [resp. $u m_2$ and $(1-u)m_2$]. Mass 
conservation implies a correlation between the sizes of the newborn particles.

%%%%%%%%%%%%%%%%%%%%%%%%%%%%%%%%%%%%%%%%%%%%%%%%%%%%%%%%%%%%%%%%%%%%%%%
\section{Kinetics}
\label{sec:kinetics}

A detailed analysis of the dynamics of the system, already at the Boltzmann 
level, according to (\ref{eq:boltzm}), is quite involved.  However, 
the key information is already present in the time evolution of the total 
density and kinetic energy densities, defined respectively as 
\begin{eqnarray}
&&n(t)=\int  f({\vec v},\sigma,t)\, d{\vec v}d\sigma 
\label{eq:dens}\\
&&E_c(t)=\int v^2\sigma^D f({\vec v},\sigma,t)\, d{\vec v}d\sigma.
\label{eq:temp}
\end{eqnarray}
The total mass density
\begin{equation}
M=\int \sigma^D f({\vec v},\sigma,t)\, d{\vec v}d\sigma
\end{equation}
is a conserved quantity and we define the temperature kinetically as
\begin{equation}
k_BT = \frac{2 E_c}{D n}  = \frac{\langle m v^2 \rangle }{D}.
\label{eq:kT}
\end{equation}
In the previous equation, the angular brackets of a quantity ${\cal Q}$ 
denote the mean value $\int {\cal Q} f({\vec v},\sigma,t)\,d{\vec v}d\sigma/n$.

From the Boltzmann equation, the time evolution for the density of particles 
reads
\begin{equation}
\frac{\D n}{\D t}=p\omega(t) n(t)
\label{eq:numbert}
\end{equation}
where $p$ is the probability that a particle breaks at a collision, and 
 $\omega(t)$  the collision frequency
\begin{equation}
\omega(t)=\frac{1}{n}\,\int d\widehat{\varepsilon}\,
d1\,d2\,\, \sigma_{12}^{D-1 }\theta(\widehat{{\vec \sigma}}
\cdot{\vec v}_{12})
\left|\widehat{\varepsilon}\cdot{\vec v}_{12}\right|f({\vec v}_1,\sigma_1,t) f({\vec v}_2,\sigma_2,t),
\label{eq:omega}
\end{equation}
where the short-hand notation $d1\,d2$ stands for 
$d{\vec v}_1 d{\sigma}_1 d{\vec v}_2 d{\sigma}_2 $.
The kinetic energy density is unaffected by the breaking events
(unlike the density itself), and only decreases due to the inelastic 
collisions:
\begin{equation}
\frac{\D E_c(t)}{\D t} \,=\, - 2b_c \gamma_0\omega(t) E_c(t).
\label{eq:tempt}
\end{equation}
In this equation,
$\gamma_0\equiv(1-\alpha^2)/(2D)$  is related to the energy lost per 
collision while $b_c$ is a dimensionless collisional
average
\begin{equation}
b_c= \frac{D}{2 \omega E_c}
\int  d{\vec{\widehat{\varepsilon}}}\, d1\,d2\,
\theta({\vec v}_{12}\cdot{\vec{\widehat{\varepsilon}}}) 
\,\left|{\vec v}_{12}\cdot{\vec{\widehat{\varepsilon}}}\right|^3 \,
\frac{m_1 m_2}{m_1+m_2} 
f({\vec v}_1,\sigma_1,t) f({\vec v}_2,\sigma_2,t),
\label{eq:bc}
\end{equation}
where it is understood
that $m_i = \sigma_i^D$. The parameter $b_c$ is a dissipation 
parameter: $\gamma_0 b_c$ is the ratio of the kinetic
energy dissipated in an ``average'' collision to the instantaneous 
mean kinetic energy per particle. 
 Eq.~(\ref{eq:numbert}) expresses the fact that due to the collisions the 
overall number of particles in the system increases in time. 
In some cases, this increase is so rapid that the 
gas exhibits a finite-time singularity,
 {\em i.e.} the number of grains become infinite in a finite amount of 
time (see next section). 
The existence of such a singularity, which can be thought of as a 
limitation of the model, signaling that some of the assumptions are too 
simple, or that some physical ingredient is missing, will 
depend on the parameters that characterize the evolution of the system,
 $p, b_c, \alpha$ and the dimensionality $D$.

%%%%%%%%%%%%%%%%%%%%%%%%%%%%%%%%%%%%%%%%%%%%%%%%%%%%%%%%%%%%%%%%%%%%%%%
\subsection{Scaling behaviour}

We will analyze the evolution of the free  evolving gas once the transient 
effects induced by the particular initial condition 
chosen have vanished. In particular, we will concentrate in the late time 
regime, addressing the issue whether the existence of a scaling solution is 
possible, where there exists a unique typical velocity and size of the 
particles characterizing the overall evolution of the system.

To this end, a good way to characterize the possible kinetic scenarios,
 is to introduce an intrinsic time scale, ${\cal C}$, which counts the 
number of collisions \cite{young}. This scale is related to time through, $\D {\cal C}=\omega(t) \D t$.
In terms of this variable, the evolution equations (\ref{eq:numbert}) 
and (\ref{eq:tempt}) read
\begin{eqnarray}
\frac{\D n}{\D {\cal C}}= p n(t)\nonumber\\
\frac{\D E_c}{\D {\cal C}}= -2 b_c\gamma_0 E_c
\end{eqnarray}
with solutions
\begin{equation}
n(t)=n_0\E^{p{\cal C}}\;,\;\;\;\;\;\;
E_c(t)=E_{c0}\E^{-2 b_c \gamma_0 {\cal C}}.
\label{eq:internalscale}
\end{equation}

Assuming the existence of a scaling regime, the kinetics of the 
system is determined by one single characteristic velocity $\bar v(t)$, and 
grain size $\bar r(t)$. The time dependence of 
the distribution function $f({\vec v},\sigma,t)$ consequently 
occurs through these two quantities [up to a trivial time dependence 
through the density resulting from the
normalization constraint (\ref{eq:dens})], and the scaling assumption amounts
to restricting $f$ to the family of functions of the form:
\begin{equation}
f({\vec v},\sigma,t) = \frac{n}{\bar r \,(\bar v)^D} \,\widetilde f({\vec c},\bar\sigma),
\label{eq:scaling}
\end{equation}
where we have introduced the rescaled size and velocity
\begin{equation}
{\vec c} = \frac{\vec v}{\bar v(t)} \quad; \quad  
\bar \sigma = \frac{\sigma}{\bar r(t)}.
\label{eq:cdef}
\end{equation}
A natural choice is to define $\bar v$ as the variance of the velocity 
distribution and $\bar r$ as the mean radius: $\bar v^2 = \langle v^2\rangle$,
$\bar r = \langle \sigma \rangle$.

Within the scaling ansatz, we obtain from (\ref{eq:bc}) that $b_c$ is time
independent, and from
(\ref{eq:omega}) that 
$\omega$ scales with time as $n \bar v \bar r^{D-1}$
(i.e. as the ratio of typical velocity over mean free path). 
If we use the 
fact that mass density is conserved (which implies that 
$\bar{r}^D n \propto M \propto t^0$), we 
know that $\bar{r}\propto 1/n^{1/D}$ and $\bar{v}^2 \propto E_c/M \propto E_c$. 
Combining these 
relations, we get for the collision frequency
\begin{equation}
\omega \propto \bar{v}n\bar{r}^{D-1} \propto \frac{\bar{v}}{\bar{r}} 
\propto n^{1/D}E_c^{1/2}
\label{eq:omeg}
\end{equation}
from which we can obtain a differential equation for the overall number of 
collisions as a function of time
\begin{equation}
\frac{\D {\cal C}}{\D  t}=\omega_0\left(\frac{n}{n_0}\right)^{\frac{1}{D}}
\sqrt{\frac{E_c}{E_{c0}}}=\omega_0 \exp\left[\left(\frac{p}{D}- b_c\gamma_0
\right){\cal C}\right]
\end{equation}
From this relation, we  can obtain the explicit time dependence of the number 
of collisions,
\begin{equation}
{\cal C}(t)=\frac{1}{-\frac{p}{D}+b_c\gamma_0}\ln\left[1+
\left(-\frac{p}{D}+b_c\gamma_0\right)\omega_0t\right]
\end{equation}
Finally, the explicit time dependence of the mean number of particles and 
kinetic energy read
\begin{eqnarray}
n(t)&=&n_0\left[1+\left(-\frac{p}{D}+ b_c\gamma_0\right)\omega_0 
t\right]^{p/\left(-\frac{p}{D}+b_c\gamma_0\right)}
\label{eq:n}\\
E_c(t)&=&E_{c0}\left[1+\left(-\frac{p}{D}+b_c\gamma_0\right)\omega_0 t
\right]^{-2b_c\gamma_0/\left(-\frac{p}{D}+b_c\gamma_0\right)}
\label{eq:ec}
\end{eqnarray}
These expressions indicate when a finite-time singularity will develop. 
Due to mass conservation, the divergence in the number 
of particles implies that the mean size of the particles vanishes.
We will see in the following sections that this feature may be
associated with a  
``shattering'' transition.
In fact, whenever the inequality
\begin{equation}
-\frac{p}{D}+ b_c\gamma_0\leq 0
\end{equation}
holds, the number of particles diverges at a time
\begin{equation}
t_f=\frac{1}{\omega_0\left[\frac{p}{D}-b_c\gamma_0\right]}
\end{equation}
at which the kinetic energy vanishes.
This inequality holds at large enough fragmentation probability,
$p\geq Db_c\gamma_0$. It therefore appears that the finite-time 
singularity cannot happen if \frag\ is just a marginal event (low $p$); alternatively, it will always be present in the  elastic limit
($\alpha=1$ so that $\gamma_0=0$).
Hence, the kinetics of a gas of freely evolving fragmenting grains is determined by 
the inelasticity and the collisional average $b_c$. In principle, this collisional average is itself a function of the inelasticity; however, as discussed in the Appendix, such a dependence only appears when the velocity dependence of the distribution function
$f({\vec v},\sigma,t)$ deviates from a Maxwellian.

In the absence of the above finite-time singularity, both $n$ and $E_c$ show
power laws with time, which, from (\ref{eq:numbert}) in turn imposes that 
$\omega \propto 1/t$. The physical meaning of this remarkable feature
(also observed in ballistic annihilation \cite{emm})
is that there exists a single relevant time scale in the problem. 
It is also noteworthy that from (\ref{eq:omeg}),
the constraint $\omega \propto 1/t$ imposes a relation between the
scaling exponents of $n$ and $E_c$.

The previous results presume the existence of a scaling regime, hypothesis 
that we have validated numerically, as will be described in Sect.~\ref{sec:numerics}.
However, note that the existence of a 
finite-time singularity does not rule out the possibility of a scaling regime. In order 
to analyze quantitatively such a regime, it is useful to 
introduce a slower time scale, $\tau$
\begin{equation}
\D\tau= n(t) \D t
\end{equation}
In this ``fake'' time scale, the finite-time singularity is no longer present. This feature favours the numerical analysis of the asymptotic scaling behaviour of the mixture. 
The internal time scale is related to $\tau$ through
 $\D {\cal C}=\omega \D \tau/n$. As a result, in terms of this new time scale, the number 
of collisions is expressed as
\begin{equation}
{\cal C}=\frac{1}{ b_c\gamma_0+\left(1-\frac{1}{D}\right)p}
\ln\left\{1+\left[b_c\gamma_0+
\left(1-\frac{1}{D}\right)p\right]\omega_0\tau\right\}
\end{equation}
 Hence, using the exponential growth of both the number of particles and 
kinetic energy in terms of the number of collisions, we get
\begin{eqnarray}
n(\tau)&=&n_0\left\{ 1+\left[b_c\gamma_0+
\left(1-\frac{1}{D}\right)p\right]\omega_0\tau
\right\}^{p/\left[b_c\gamma_0+\left(1-\frac{1}{D}\right)p\right]}
\label{eq:ntaugeneral}\\
E_c(\tau)&=&E_{c0}\left\{ 1+\left[b_c\gamma_0+\left(1-\frac{1}{D}
\right)p\right]\omega_0\tau\right\}^{-2D/\left[D+\frac{(D-1)p}{b_c\gamma_0}\right]}
\label{eq:entaugeneral}
\end{eqnarray}
When compared with (\ref{eq:n}) and  (\ref{eq:ec}), one can see that, although qualitatively similar, the additional factors that characterize the exponents in this time scale ensure the absence of a finite-time singularity.

The time evolution of the physical quantities of interest in the scaling regime, as depicted in (\ref{eq:ntaugeneral}) and  (\ref{eq:entaugeneral}),
depend on the inelasticity coefficient, $\alpha$, the fragmentation probability, $p$, 
the dimensionality, $D$, and the collisional average $b_c$. From all of these, only $b_c$ 
is not known {\it a priori}. In the Appendix, we show how it is 
related to the particle distribution. In particular, it is a constant for a 
Maxwellian distribution of velocities with a single mass-independent temperature. 
In this limiting case, analytic expressions for 
the exponents can be obtained. However, as we will 
see, such a factorization is too simplistic in generic situations.
The collisional average will therefore have to  be determined numerically, and in all 
cases it is the only  input required to predict the  corresponding 
exponents governing the scaling 
behaviour of the freely evolving gas.

Incidentally, the existence of  a scaling regime can be used at one's 
advantage in DSMC. 
As has been explained previously, one of the numerical 
difficulties in studying 
the fragmentation of a freely evolving granular gas lies in the fact that the 
collision 
frequency decreases with time, making it harder to work with larger systems sizes 
(which is what inevitably happens in this case due to he fragmentation). 
On the contrary, working with the rescaled velocities ${\vec c}$ instead
of the original ${\vec v}$ significantly speeds up the program, and allows
to probe longer time scales \cite{Montanero}.

%%%%%%%%%%%%%%%%%%%%%%%%%%%%%%%%%%%%%%%%%%%%%%%%%%%%%%%%%%%%%%%%%%%%%%%
\subsection{Kinetic scenarios}

As an illustration of the various  kinetic scenarios 
covered by the previous analysis, let us 
consider in more detail a few limiting cases related to the kinetics of freely 
evolving gases.

In the absence of fragmentation, $p=0$, obviously the number of particles remains 
constant, the system corresponds to that of a monodisperse 
granular gas, and the energy decays as
\begin{equation}
E_c=\frac{E_{c0}}{\left[1+b_c\gamma_0\omega_0 t\right]^2}
\end{equation}
recovering Haff's law \cite{brito}. Here, the collisional parameter
$b_c$ is close to 1, with deviations originating from the non-Gaussian
features of the velocity distribution \cite{vannoije}. In this particular case, 
the two time scales $\tau$ and $t$ are just proportional to each other, 
yielding the same exponent. 
The assumption of spatial homogeneity prevents us 
from addressing the late stage deviations from Haff's law that have been 
described in freely evolving granular gases \cite{brito}. 

For an elastic gas, $\alpha=1$, the kinetic energy is conserved. If grains can still 
fragment with a probability $p$, the number of particles will always 
diverge at a finite time,
\begin{equation}
n(t)=\frac{n_0}{\left[1-\frac{p}{D}\omega_0t\right]^D}
\label{eq:nt}
\end{equation}
In this case it is easier to appreciate the advantage of introducing the 
slower time scale $\tau$. In terms of this variable, the number of particles 
will increase as
\begin{equation}
n(\tau)=n_0\left[1+\left(1-\frac{1}{D}\right)p\omega_0\tau\right]^{D/(D-1)}
\label{eq:ntau}
\end{equation}
In order to analyze the existence of an asymptotic regime (hypothesis upon 
which both (\ref{eq:nt}) and (\ref{eq:ntau}) have been derived), it is 
more convenient to use the latter time scale, because it does not lead to a finite-time divergence. 
In this way, it is possible to make use of the standard techniques to address the  
appearance of power-law kinetic regimes, since the indefinite growth ensures the possibility 
to span several decades. Hence,  
once the algebraic regime is achieved, accurate values of the exponents can be derived as well. If such exponents are characterized,  it is always possible to obtain their counterparts 
controlling the evolutions in 
the real time scale $t$.

The technique of resorting to a slower time scale will always be useful for the present 
class of fragmentation problems, because the  exponent appearing in 
 (\ref{eq:ntaugeneral}), characterizing the increase in the grain number density, 
is always positive for any dimension larger than 1. For $D=1$, the elastic limit 
is singular. In this case, the scale ${\cal C}$ is 
proportional to $\tau$ and the number of particles increases 
exponentially in $\tau$ rather than algebraically. This behavior is recovered, 
by performing the limit $D\rightarrow 1$ in (\ref{eq:ntaugeneral}), 
and is due to the fact that 
in one dimension the collision frequency is proportional 
to the number of particles for our elastic mixture.

%%%%%%%%%%%%%%%%%%%%%%%%%%%%%%%%%%%%%%%%%%%%%%%%%%%%%%%%%%%%%%%%%%%%%%%
\section{Numerical simulations}
\label{sec:numerics}

%%%%%%%%%%%%%%%%%%%%%%%%%%%%%%%%%%%%%%%%%
\subsection{The method}

We have implemented the Direct Simulation 
Monte Carlo technique (DSMC), which offers a particle method to solve 
numerically
a Boltzmann-type equation, such as (\ref{eq:boltzm}). 
This method was originally introduced to study 
rarefied gases \cite{bird}, and has been extended to study granular 
gases \cite{santos}. It has been 
used in a number of different situations and it is easy to adapt to the 
present system, where the number of grains increases with time.
There exist already a number of applications to situations where the number 
of particles is not a conserved quantity, as it is the case in ballistic 
annihilation \cite{emm}. 

The assumption of spatial homogeneity simplifies further the implementation 
of DSMC, because it is enough to sample uniformly the relative orientation of 
the colliding pair, $\widehat{\varepsilon}$, without the need to keep track of the 
positions of the particles. Sequentially, a pair of particles is selected at 
random. Also the vector $\widehat\varepsilon$ is chosen randomly on the unit sphere 
and  the kinematic constraint is checked. In case the particles can collide, 
the cross section is computed and the collision takes place proportionally 
to its value. The decrease in velocity of the particles due to the 
inelasticity reduces the cross section, hence the simulations become 
more expensive as time evolves. After the collision is executed (according 
to (\ref{eq:collision})), one of the two particles is chosen randomly, 
and it is broken with probability $p$. 

The collision frequency of each collision $i$-$j$ can be computed on the basis of 
the cross-section, and hence the time is advanced accordingly at each 
collision
\begin{equation}
t \to t + \frac{1}{n^2 \left|v_{ij}\cdot\widehat \varepsilon\right| \,\sigma_{ij}^{D-1}},
\end{equation}
where $n$ is the instantaneous number density.
It is then possible to keep track of the kinetic evolution of 
the system, as well as gather statistics on the distribution of sizes and 
velocities. In the following section we will analyze both the kinetics and 
the -- self-similar in time-- polydisperse distributions that 
the free evolution of the gas 
gives rise to.

%%%%%%%%%%%%%%%%%%%%%%%%%%%%%%%%%%%%%%%%%%%%
\subsection{Numerical results: Kinetics}

We have first analyzed the kinetics in the homogeneous regime to 
test the scaling hypothesis discussed in Sect.~\ref{sec:kinetics}.
To this end, we have used DSMC simulations 
to study the evolution of a two-dimensional granular gas 
of circular disks with constant restitution coefficient $\alpha$.
 In all cases we start from an initial condition where all particles have the same size 
and their velocities are distributed according to a Maxwellian 
(monodisperse gas of elastic disks). As it has 
already been described in Sect.~\ref{sec:model}, we restrict ourselves to the 
simplest model where colliding particles fragment with a uniform probability 
independent of the collision properties, and where only one of the post-colliding 
disks fragments in two out-coming grains.

In order to validate the existence of a scaling regime, we have computed the 
total number of particles, $n(t)$, the collision frequency, $\omega (t)$, the mean velocity, 
$\langle v^2\rangle^{1/2}$, as well as the mean radius, $\langle \sigma \rangle=\bar{r}$.
 We have also computed different moments of the size, $\langle \sigma^n\rangle(t)$, and 
velocity, $\langle v^2\rangle^{n/2}(t)$, distribution functions. 
Within the scaling picture, we expect $\langle \sigma^n \rangle \propto (\bar r)^n$
and $\langle \left|v\right|^p\rangle \propto (\bar v)^p$, as far 
as the time dependence is concerned. The possibility of a multi-scaling behaviour 
has been reviewed by Ben-Naim and Krapivsky \cite{BenNaim}.

In Figure~\ref{fig:generalalpha0}a, we display the time evolution of a few relevant 
quantities in the real time scale, $t$. For the parameters chosen, the system shows a 
finite-time singularity. As already indicated, such a phenomenon makes it 
difficult to assess the existence of a scaling regime, and complicates its
analysis. On the contrary, when displayed 
in terms of the slower time scale $\tau$, as shown in Fig.~\ref{fig:generalalpha0}b, 
one can see the existence of a scaling regime, which can be characterized.
An alternative possibility to assess the existence of a scaling regime
without introducing $\tau$ would be to check if $\langle \sigma^p\rangle$
scales like $\langle \sigma^p\rangle^{n/p}$. 

In  Figure~\ref{fig:taualpha0}a 
we display different moments of the velocity 
distribution, as well as 
the mean collisional velocity $\langle v_{\hbox{\scriptsize coll}}\rangle$, 
defined as the mean velocity modulus of colliding
partners, i.e restricting to those particles that are in precollisional
configurations (since a typical collision involves a particle that is ``hotter''
than the mean background, this quantity is larger than $\langle v \rangle$,
as can be observed on the figure). 
Due to the inelasticity of the collisions, the 
typical velocity of the particles decays as a function of time. The figures 
show that such a decay is algebraic in $\tau$, and also that there is a single 
exponent characterizing different moments of the velocity. 
The same holds for the characteristic size of the particles 
as a function of time, as depicted in Fig.\ref{fig:taualpha0}b. 
In this case however, the scaling does not 
hold for negative powers of the size. We will see 
in the next section how such a deviation from simple scaling 
relates to the peculiarities of the size distribution, 
but it does not imply a breaking  of the scaling hypothesis; it only 
signals the fact that moments of negative order 
are controlled by the smallest species in the mixture.

We can characterize the asymptotic regime by a set of exponents, that in the 
terms of $\tau$ can be written generically as
\begin{eqnarray}
n(\tau)&\sim&\tau^{\varepsilon '}\nonumber\\
E(\tau)&\sim&\tau^{-2 \gamma '}\nonumber\\
\sigma(\tau)&\sim&\tau^{-\beta '}.
\label{eq:expo}
\end{eqnarray}
Once these exponents are known, their counterparts associated with the $t$ scale
follow. For example, using that $\D \tau=n_0(1+a\tau)^{\varepsilon'}\D t $,  
for $\varepsilon'\neq 1$ we have 
\begin{equation}
\frac{1+a\tau}{1+a\tau_0}=\left[1+(1-\varepsilon')\frac{n_0a(t-t_0)}{(1+a\tau_0)^{1-\varepsilon'}}\right]^{1/(1-\varepsilon')}
\end{equation}
from which we can derive the time evolution of any physical quantity.

In Table \ref{table:expon} we give the values of the exponents fitted 
numerically for the 2-dimensional gas at different values of the inelasticty 
parameter. We can determine the different exponents independently, but they 
are related to each other since all of them are eventually functions of the 
collisional average $b_c$. It is in fact rather straightforward to check 
that, once one of the three exponents is known, the remaining ones can be determined 
using the relationship between them as described in Sect.~~\ref{sec:kinetics}.

\begin{table}
\caption{Exponents characterizing the time evolution of 
the number and energy densities as well as the mean particle radius (see 
(\ref{eq:expo}) for definitions) for different values of the inelasticity 
parameter $\alpha$, when the probability of grain fragmentation at a collision 
is $p=1/2$}
\begin{center}
\renewcommand{\arraystretch}{1.4}
\setlength\tabcolsep{15pt}
%\begin{tabular}{@{}llp{1.8cm}l}
\begin{tabular}{lllll}
\hline\noalign{\smallskip}
$\alpha$ & $\varepsilon '$ & $\beta '$ & $\gamma '$ \\
\noalign{\smallskip}
\hline
\noalign{\smallskip}
0.3 & 0.95 & 0.45 & 0.55  \\
0.5 & 1.05 & 0.5  & 0.475 \\
0.7 & 1.3  & 0.65 & 0.35  \\
0.9 & 1.9  & 0.95 & 0.125 \\
0.95& 2.3  & 1.2  & 0.05  \\
\noalign{\smallskip}
\hline
\noalign{\smallskip}
\end{tabular}
\end{center}
\label{table:expon}
\end{table}

In Table \ref{tab:bc1} we list the values of the collisional average $b_c$  
computed from the numerical simulations. We also display the values of the 
exponent $\varepsilon '$ derived using the corresponding computed values 
of $b_c$; comparing such values with those measured 
numerically (as listed in Table~\ref{table:expon}) shows a good agreement, 
illustrating the consistency of the analysis.
The scaling hypothesis also implies, from (\ref{eq:internalscale}),
that
\begin{equation}
E_c \propto \left(\frac{n}{n_0}\right)^{2 b_c \gamma_0/p}
\label{eq:ecexp}
\end{equation}
which provides an alternative way to 
compute $b_c$ in the scaling regime (where it has to be time independent). We 
display in Table~\ref{tab:bc1} the collisional average computed following
this route [under the symbol $b_c^{(1)}$]. It shows a good agreement with the 
computed collisional average, giving additional support to the existence 
of an asymptotic scaling regime.

In the Appendix, we have given a generic expression for $b_c$, showing 
that it depends sensitively on both the size and the velocity distribution
of the mixture in the scaling regime. If the velocity dependence enters only through 
a Maxwellian, then the value of $b_c$ is independent of the mass distribution. This 
surprising result shows that this collisional average is a quantity that will be 
sensitive to deviations from Maxwell distributions. In this contribution we have 
only analyzed the behavior of the gas within the molecular chaos 
assumption, and 
hence, deviations of $b_c$ from unity signal a deviation from Maxwellian behavior
(with the strong assumption that the temperature of a given species does
not depend on its size). 
In  a more generic study, deviations from Maxwellian behavior may also
result from the 
breakdown of molecular chaos, so that the use of $b_c$ has to be supplemented by a 
second quantity that is sensitive to such a breakdown (e.g. the averaged 
impact parameter \cite{luding,presand}).

For a  mono-disperse granular gas, deviations from Maxwellian behavior vanish 
as the elastic limit is approached and can be described 
quite accurately in terms of a Sonine expansion 
around the Maxwellian \cite{vannoije}. The values of $b_c$ displayed in 
Table~\ref{tab:bc1}, on the contrary, show that the deviations  
from a Maxwellian velocity distribution become more pronounced
as one approaches the elastic limit. 
This counter-intuitive trend can be traced back to the peculiar asymptotic size distribution of 
particles close to elasticity. Such a behavior is described in detail in the next section
where it appears that 
the velocity distribution for nearly elastic gases deviates strongly from a 
global Maxwellian. At higher inelasticities, such deviations, even if always present, 
are not so dramatic.

\begin{table}
\caption{Collisional average $b_c$, and predicted 
exponent $\varepsilon '$  for different values of the 
inelasticity parameter $\alpha$, when the breaking probability 
is $p=1/2$. $b_c^{(1)}$ is obtained from 
(\ref{eq:ecexp}), plotting the kinetic energy as a function 
of the number of particles.}
\begin{center}
\renewcommand{\arraystretch}{1.4}
\setlength\tabcolsep{15pt}
%\begin{tabular}{@{}llp{1.8cm}l}
\begin{tabular}{llll}
\hline\noalign{\smallskip}
$\alpha$ & $b_c$ & $b_c^{(1)}$ & $\varepsilon_{\mbox{th}}'$\\
\noalign{\smallskip}
\hline
\noalign{\smallskip}
0.3 & 1.24 & 1.248  &   0.94   \\
0.5 & 1.18 & 1.173 &   1.06 \\
0.7 & 1.07 & 1.042 &   1.29  \\
0.9 & 0.67 & 0.645 &  1.77 \\
0.95& 0.39 & 0.2318 &  1.94 \\ 
\noalign{\smallskip}
\hline
\noalign{\smallskip}
\end{tabular}
\end{center}
\label{tab:bc1}
\end{table}

%%%%%%%%%%%%%%%%%%%%%%%%%%%%%%%%%%%%%%%%%%%%%%%%%%%%%%%%%%%%%%%%%%%%%%%
\subsection{Numerical results: Distribution functions}

The existence of a scaling regime implies that the distribution function
of a given quantity at two different times has to be invariant
after an appropriate rescaling of this quantity [see (\ref{eq:scaling})
and (\ref{eq:cdef})]. For instance, the distribution of 
$\bar \sigma = \sigma/\langle\sigma\rangle$ is expected to be time independent.
 From the DSMC simulation runs, we have computed scaled distribution 
functions, which do indeed reach a steady state, in agreement with the 
scenario developed in Sect.~\ref{sec:kinetics}. We will now discuss 
in detail the characteristic features of some of the relevant 
stationary distributions.

%%%%%%%%%%%%%%%%%%%%%%%%%%%%%%%%%%%%%%%%%%%%%%%%%%%%%%%%%%%%%%%%%%%%%%%
\subsubsection{Size distribution}

The fragmentation process generates a polydisperse mixture of grains that evolves 
continuously in time, with a decreasing mean size. 
Large particles are consequently progressively destroyed, but the shape of the 
size distribution is preserved, in such a way that the relative amount 
of particle with respect to the time dependent mean size, $\langle \sigma \rangle  (t)$,
 becomes independent of time. 

In this regime, the scaled size distribution exhibits a marked 
peak for vanishingly small grains, indicating that smaller particles become 
predominant in the mixture. 
In Figure \ref{fig:histn} we show such distributions for a number of 
inelasticity values, where two regions can be identified. On the 
extreme of large grains, the decay is exponential while 
it is algebraic for small particles.
The exponent characterizing 
this algebraic behavior increases with $\alpha$. For example, for
$\alpha=0.3$ the  exponent is consistent with $P(\sigma)\sim \sigma^{-0.1}$, 
while for  $\alpha=0.7$ we can fit the curve numerically 
with $P(\sigma)\sim \sigma^{-0.6}$.

It is interesting to note that 
at $\alpha=0.9$, and also at $\alpha=0.95$,  the above exponent becomes close to
 $-1$. Such a value is peculiar, because it implies that  the probability 
distribution is not integrable at the origin, incompatible with the 
requirement that $P(\sigma)$ [or $P(\bar\sigma)$] has to be normalizable.
In fact, a careful inspection of  Fig.~\ref{fig:histn} shows 
that  $P(\sigma)$  at high inelasticity develops a jump very 
close to the origin. Such a jump is consistent with the  appearance
of a singular contribution, due to a finite 
fraction of particles with vanishing size.  This behaviour is reminiscent
of a ``shattering'' transition \cite{redner} and is corroborated 
by the velocity distributions described in the next section.
It indicates the development 
of an intrinsic size heterogeneity in the system.

The algebraic decay of $P(\sigma)$ implies that, even when the exponent is 
smaller (in absolute value)
than $-1$, certain averages of the distribution function will not exist. 
This is consistent with the behavior described in the previous section, and simply 
indicates that in general, averages of negative powers of the grain size will 
be controlled by the smallest species available in the mixture, rather than by the 
overall mixture. However, such a lack of scaling for certain moments of the 
distribution is not incompatible with the dynamic scaling behavior of the complete 
distribution function; rather, it shows some of the peculiarities of the 
asymptotic behavior of the scaling distribution (in this case at small sizes).

The existence of a stationary size distribution suggests the possibility to map the 
fragmenting granular gas onto an effective polydisperse mixture with a time dependent 
mean. Although this is indeed possible in principle, we do not have any theoretical 
framework (except the complicated initial Boltzmann equation) to predict
either the algebraic decay of $P(\sigma)$  or the details of 
the exponential tail. 

%%%%%%%%%%%%%%%%%%%%%%%%%%%%%%%%%%%%%%%%%%%%%%%%%%%%%%%%%%%%%%%%%%%%%%%
\subsubsection{Velocity distribution}

We have also studied the velocity distribution  to assess how 
relevant deviations from Maxwellian are, and also to analyze the 
meaning of a global temperature characterizing the whole system.

In Figure~\ref{fig:pvxrandom} we display the distribution 
of rescaled velocities  ${\vec c}$ at different 
inelasticities. Deviations from 
Maxwellian behavior are observed at all values of $\alpha$. However, the 
deviations become more prominent close to the elastic limit, in agreement 
with the development of a singular contribution coming from vanishingly 
small grains, explaining the small values for $b_c$ listed in Table~\ref{tab:bc1}. 
 In order to gain more insight, we have also computed 
separately the velocity distribution from those particles with a 
mass smaller than $5\%$ the mean mass. In Figure~\ref{fig:pvxspecies95}a
we compare the overall velocity distribution  with 
that of the smaller particles and the 
contribution coming from the rest of the grains for $\alpha=0.95$. 
The two subsets behave more closely to a Maxwellian than 
the mean velocity distribution. The low velocity part of the 
distribution is controlled by the contribution from most of 
the particles, while the large velocity part is dominated by the small 
particles. These are characterized by a much larger temperature 
than the mean, while the rest of the species are much colder 
(large particle have suffered a low number of collisions (and hence
of fragmentations), which requires that they were among the very ``cold'' ones initially). 

Although less dramatic, the same trend is observed for all values of $\alpha$. 
For example, in Fig.\ref{fig:pvxspecies95}b we display the equivalent 
distribution functions at $\alpha=0.5$. In this case, again the two 
subsets of grains evolve following separate approximate Maxwellians. 
According to this interpretation, as $\alpha$ increases, the separation 
in temperates between the two sets of particles increases. 

Such an inhomogeneous dependence of the velocity distribution (and in 
particular its second moment) on grain species, explains the  large deviations of
 $b_c$ from the Maxwellian prediction reported close to the elastic limit.
These results also cast doubts on the use of a single temperature to 
characterize a polydisperse granular mixture. Nonetheless, it is worth 
noting that such a size heterogeneity does not affect the scaling hypothesis. 
Even if different temperatures could be defined for different 
subsets of particles, all of them have the same time evolution, explaining 
why it is possible to have a simple velocity scaling. Such a behavior has been 
reported for the free evolution of bidisperse granular 
mixtures \cite{garzo}. However, the absence of a common temperature 
invalidates the use of perturbative calculations, e.g. through a Sonine 
expansion, to determine the deviations from Gaussian behavior as the 
inelasticity is increased, as is customary in the case of monodisperse 
(and even bidisperse) granular gases.

%%%%%%%%%%%%%%%%%%%%%%%%%%%%%%%%%%%%%%%%%%%%%%%%%%%%%%%%%%%%%%%%%%%%%%%
\section{Discussion}

We have investigated  the evolution of a granular gas in the absence 
of any driving, taking into account the possibility that the collisions
may lead to a \frag\ of the impacting grains.

Rather than performing a detailed analysis of the fragmentation process, 
we have analyzed the simplest possible model, where the fragmentation 
is a random event uncorrelated to the details of the collision, except for the 
requirement that only colliding particles may fragment. This simple fact already 
leads to a number of interesting results. We have shown the existence of a 
scaling homogeneous regime characterizing the kinetics of the system. Such 
a regime is controlled eventually by a single collision average, $b_c$. 
The relevance of $b_c$ in the asymptotic regime shows 
that a  theory based on rate equations, usually disregarding collision induced
correlations,  will be too simplistic to capture the dynamics in these systems.

We have shown how the evolution in the scaling regime may lead to a finite-time 
singularity where the number of particles diverges. 
We have derived the conditions leading to this finite-time catastrophe, 
depending on the different 
parameters that characterize the system. 
In the absence of such a singularity, the various moments 
of the size and velocity distribution exhibit algebraic scaling
laws with time, that can be regarded as a
generalized Haff's law. 

The size distribution exhibits two different limiting kinds of behaviour:
an exponential tail at large sizes (compared to the time dependent 
mean size), and a 
power-law distribution for small sizes, which 
may develop into delta singularity close to the elastic limit. 
The details and character of such a transition, which
has been  reported in other fragmenting systems \cite{redner}
deserve a more careful analysis.
The existence of a scaling regime is not incompatible with the presence of marked 
size heterogeneities in the mixture. In particular, the characteristic temperature of 
small and large grains can be significantly different, in such a way that the mean 
scaled velocity distribution  cannot be regarded as a slightly distorted Maxwellian.

In order to assess to which extent the results discussed are specific of 
the fragmentation mechanism that has been chosen, we have also carried out numerical 
studies for a model where the fragmentation probability is correlated to the 
energy lost in a collision \cite{prep}. In that case there exists still a scaling 
regime characterizing the kinetics. The scaled distribution functions are also 
similar, except for the fact that size distributions are closer to an exponential. 
Hence, the results obtained through the analysis of the simplest model provide a 
reference system, against which other models can be  tested, and in this way it 
should be possible to disentangle the generic aspects involved in the dynamics 
of fragmenting granular materials.

\medskip  
Acknowledgments: We would like to thank A. Barrat
for useful discussions. I.P. acknowledges the support  of 
CNRS through the attribution of a ``poste de chercheur associ\'e'',
and the hospitality of the ``Laboratoire de Physique Th\'eorique'' in Orsay
where most of this work was performed.

\appendix
\section*{Appendix: On the collisional average $b_c$}
\label{sec:app}
The kinetics of the free fragmenting granular gas depends on the value 
of the collisional average $b_c$, which in turn is a complicated 
function of the joint size and velocity distribution.
To proceed further, we assume that in the scaling regime, the velocity dependence 
is Maxwellian. From the distribution function $f({\vec v}, \sigma,t)$, we 
change variables and introduce the joint distribution of masses and velocity
$f(m,{\vec v},t)$ that we approximate by
\begin{equation}
f(m_i,{\vec v}_i,t)\simeq\psi(m_i,t) \phi(m_iv_i^2,t)
\setcounter{eqsubcnt}{0}
\end{equation}
where $\psi$ is the mass distribution, while  $\phi(m_iv_i^2) $
is the Maxwellian,
\begin{equation}
\phi(m_iv_i^2,t)=\left(\frac{\beta m_i}{2\pi}\right)^{D/2}\E^{-m_iv_i^2\beta/2}.
\end{equation}
The time dependence of $\phi$ is
encoded in $\beta = 1/(k_BT)$, see (\ref{eq:kT}). 
 The key assumption here is 
that $k_BT$ does not depend on the mass $m$, as would be the case in an equilibrium
polydisperse system. We have
\begin{equation}
b_c=\beta\frac{\int \D 1\,\D 2\,
\D \widehat{\varepsilon}\,\theta({\vec v}_{12}\cdot\widehat{\varepsilon})\left|{\vec v}_{12}
\cdot\widehat{\varepsilon}\right|^3\sigma_{12}^{D-1}\umu_{12}\psi(m_1) \phi(m_1v_1^2)\psi(m_2) 
\phi(m_2v_2^2)}{\int \D 1\,\D 2\,\D \widehat{\varepsilon}\,
\theta({\vec v}_{12}\cdot\widehat{\varepsilon})\sigma_{12}^{D-1}\psi(m_1) \phi(m_1v_1^2)\psi(m_2)
 \phi(m_2v_2^2)}
\end{equation}
where $\D 1$ stands for $\D {\vec v}_1\,\D m_1$.
We can express the collisional average $b_c$ in dimensionless quantities, 
and also transform the velocities to express them in terms of the 
center-of-mass and relative coordinates. One arrives at
\begin{eqnarray}
b_c&=&\frac{2\beta }{(D+1)}\frac{\int \D {\vec c}\D m_1\D m_2c^3 \sigma_{12}^{D-1}\umu_{12}
\psi(m_1)\psi(m_2)(m_1m_2)^{D/2}\E^{-\beta\umu_{12} c^2/2}}{\int
 \D {\vec c}\D m_1\D m_2c\sigma_{12}^{D-1}\psi(m_1)\psi(m_2)(m_1m_2)^{D/2}
\E^{-\beta\umu_{12}c^2/2}}\nonumber\\
&=&\frac{\int \D m_1\D m_2\left(
\frac{1}{\umu_{12}}\right)^{\frac{D+3}{2}}\umu_{12}\,\sigma_{12}^{D-1}\,\psi(m_1)
\psi(m_2)(m_1m_2)^{D/2}}{\int \D m_1\D m_2 \left(
\frac{1}{\umu_{12}}\right)^{\frac{D+1}{2}}\,\sigma_{12}^{D-1}\,\psi(m_1)
\psi(m_2)(m_1m_2)^{D/2}}=1
\end{eqnarray}
where use has been made of
\begin{equation}
\int \D \widehat{\varepsilon}\theta({\vec c}\cdot\widehat{\varepsilon})\left|{\vec c}
\cdot\widehat{\varepsilon}\right|^n=\frac{\pi^{\frac{D-1}{2}}\Gamma
\left(\frac{n+1}{2}\right)}{\Gamma\left(\frac{n+D}{2}\right)} \, c^n
\end{equation}
So, $b_c=1$ if the velocity distribution follows a Maxwellian. Hence, this 
collisional average is a good measure of deviations from Maxwellian behavior 
in polydisperse granular gases.

\bibliography{fragment_new}

\newpage
\section*{Figure Captions}
\begin{figure}
\caption[]{Time evolution in linear-log scale of the total number of particles, 
 mean size and 
characteristic velocities for $\alpha=0.5$, and a fragmentation probability
 $p=1/2$ in $D=2$. {\bf (a)} Time evolution as a function of the real time scale $t$; {\bf (b)} Time evolution in a log-log scale in terms of the slow time scale $\tau$. The thin lines correspond to the fitted power-laws for the different quantities.}
\label{fig:generalalpha0}
\end{figure}
\vspace{1cm}
\begin{figure}
\caption[]{Time evolution of different moments of the {\bf (a)} velocity  
distribution function, and {\bf (b)} size distribution function for the same parameters as Fig.~\ref{fig:generalalpha0}. In {\bf (a)} 
the time evolution of the mean velocity of colliding grains is also depicted. 
In both cases the slow time scale is used, and the corresponding fitted power-law decays are displayed as thin lines.}
\label{fig:taualpha0}
\end{figure}
\vspace{1cm}
\begin{figure}
\caption[]{Log-log plot of the rescaled particle radius distribution, 
for a uniform random fragmentation probability $p=1/2$ in  $D=2$.}
\label{fig:histn}
\end{figure}
\vspace{1cm}
\begin{figure}
\caption[]{Linear-log plot of the rescaled velocity distribution
${\bf c} = {\bf v}/\bar v$.}
\label{fig:pvxrandom}
\end{figure}
\vspace{1cm}
\begin{figure}
\caption[]{Linear-log plot of particle velocity distributions, scaled by 
the mean velocity, for a uniform fragmentation probability. ``Small'' refers to the distribution of 
velocities for particles having a mass smaller than 5\% the mean mass. ``Large'' 
corresponds to the distribution of the rest of the grains. {\bf (a)} $\alpha=0.95$; 
{\bf (b)} $\alpha=0.5$ }
\label{fig:pvxspecies95}
\end{figure}
\vspace{1cm}
\newpage
\section*{Figures}
\begin{figure}
\begin{center}
\includegraphics[width=0.75\textwidth]{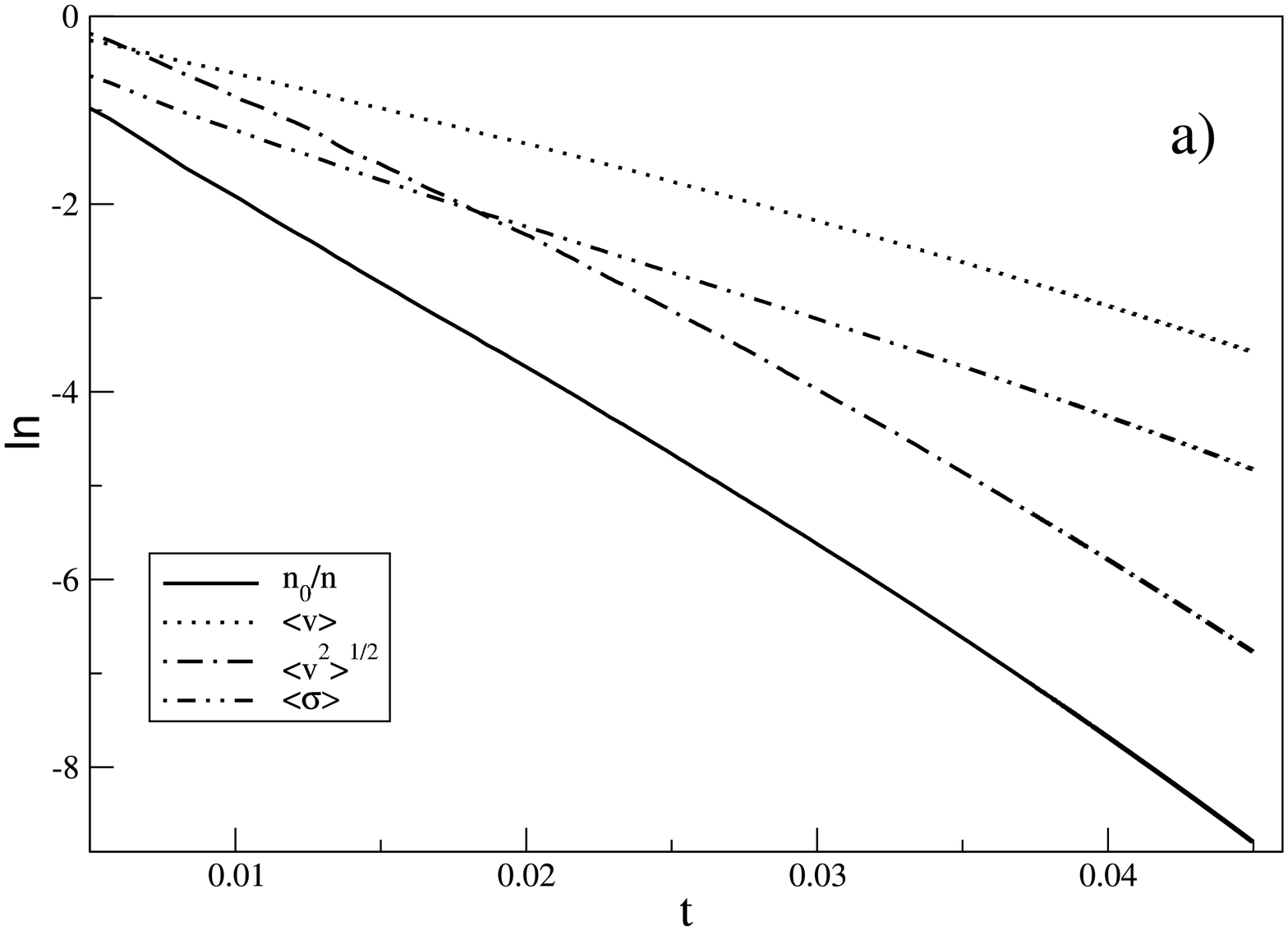}
\end{center}
\end{figure}
\begin{figure}
\begin{center}
\includegraphics[width=0.75\textwidth]{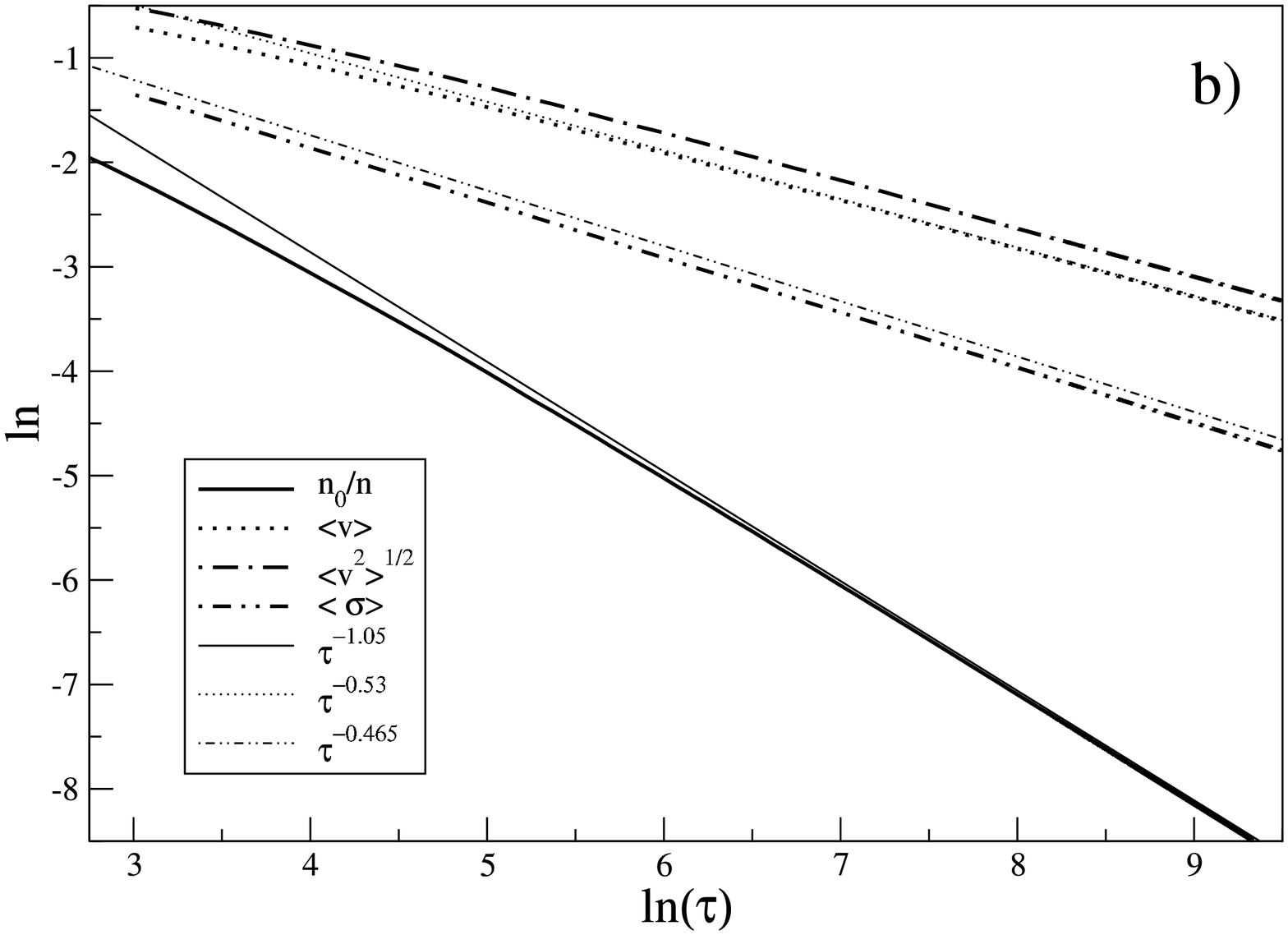}
\end{center}
\end{figure}
\vspace{1cm}
\newpage
\begin{figure}
\begin{center}
\includegraphics[width=0.75\textwidth]{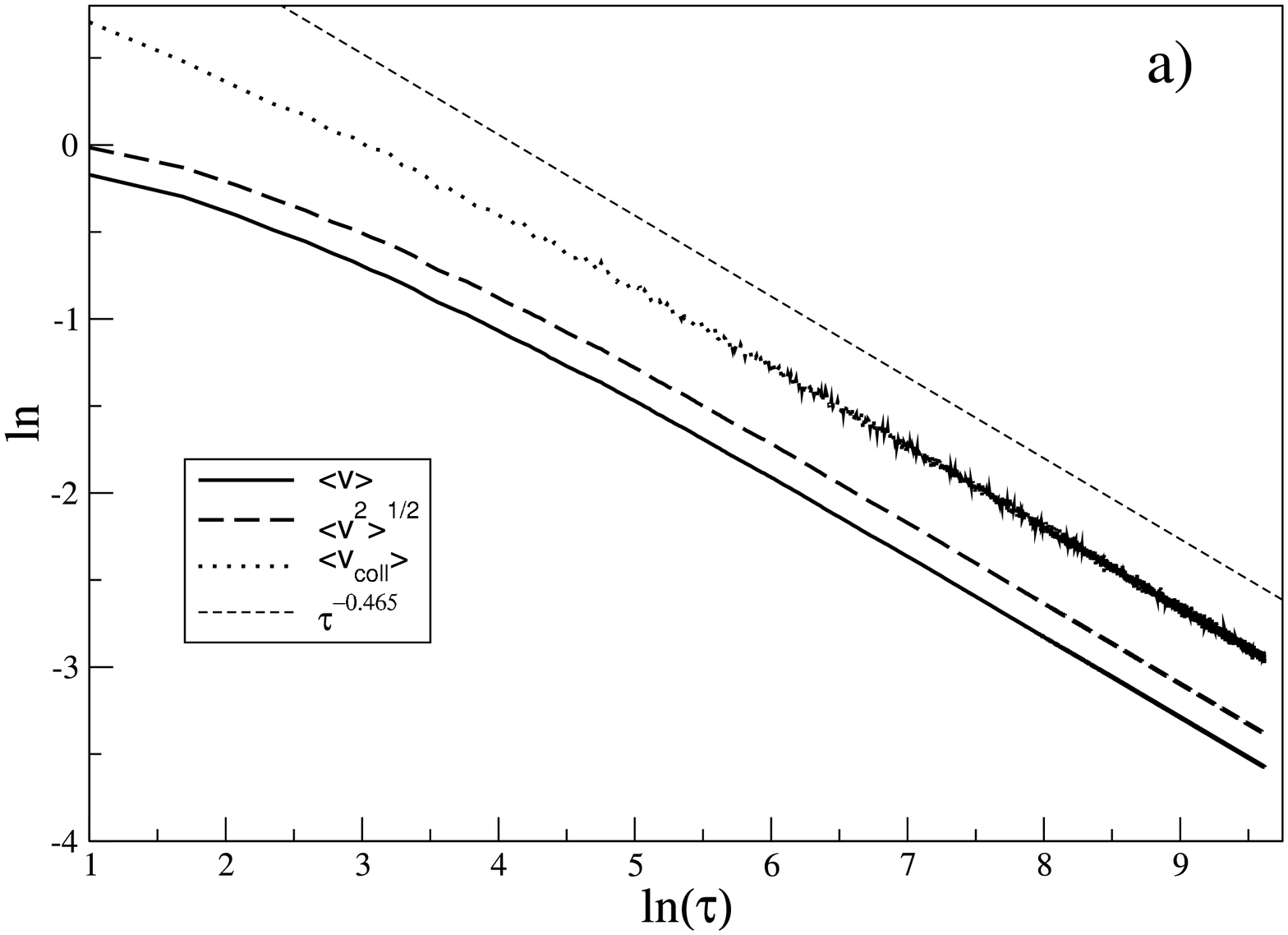}
\end{center}
\end{figure}
\begin{figure}
\begin{center}
\includegraphics[width=0.75\textwidth]{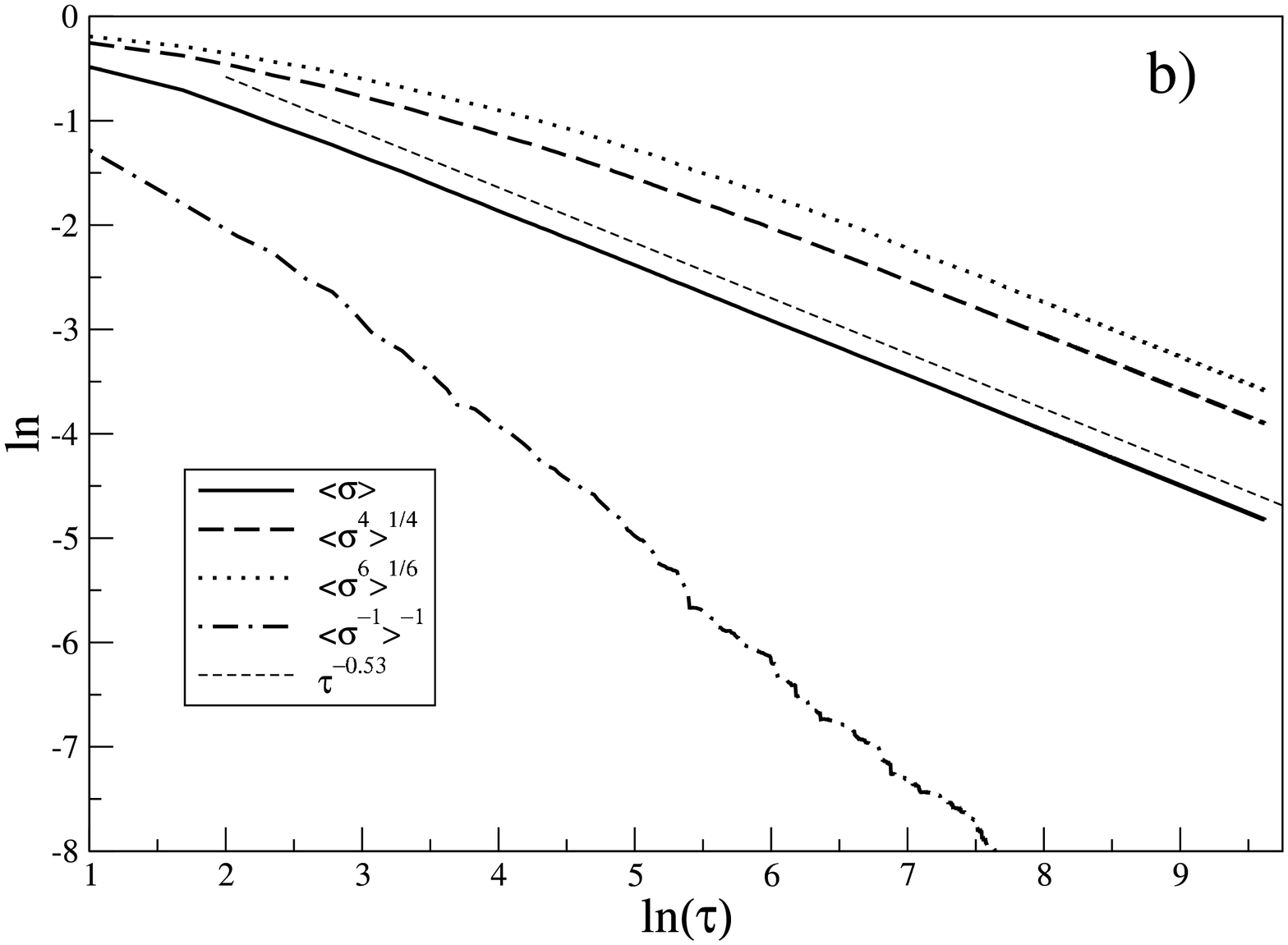}
\end{center}
\end{figure}
\vspace{1cm}
\newpage
\begin{figure}
\begin{center}
\includegraphics[width=.75\textwidth]{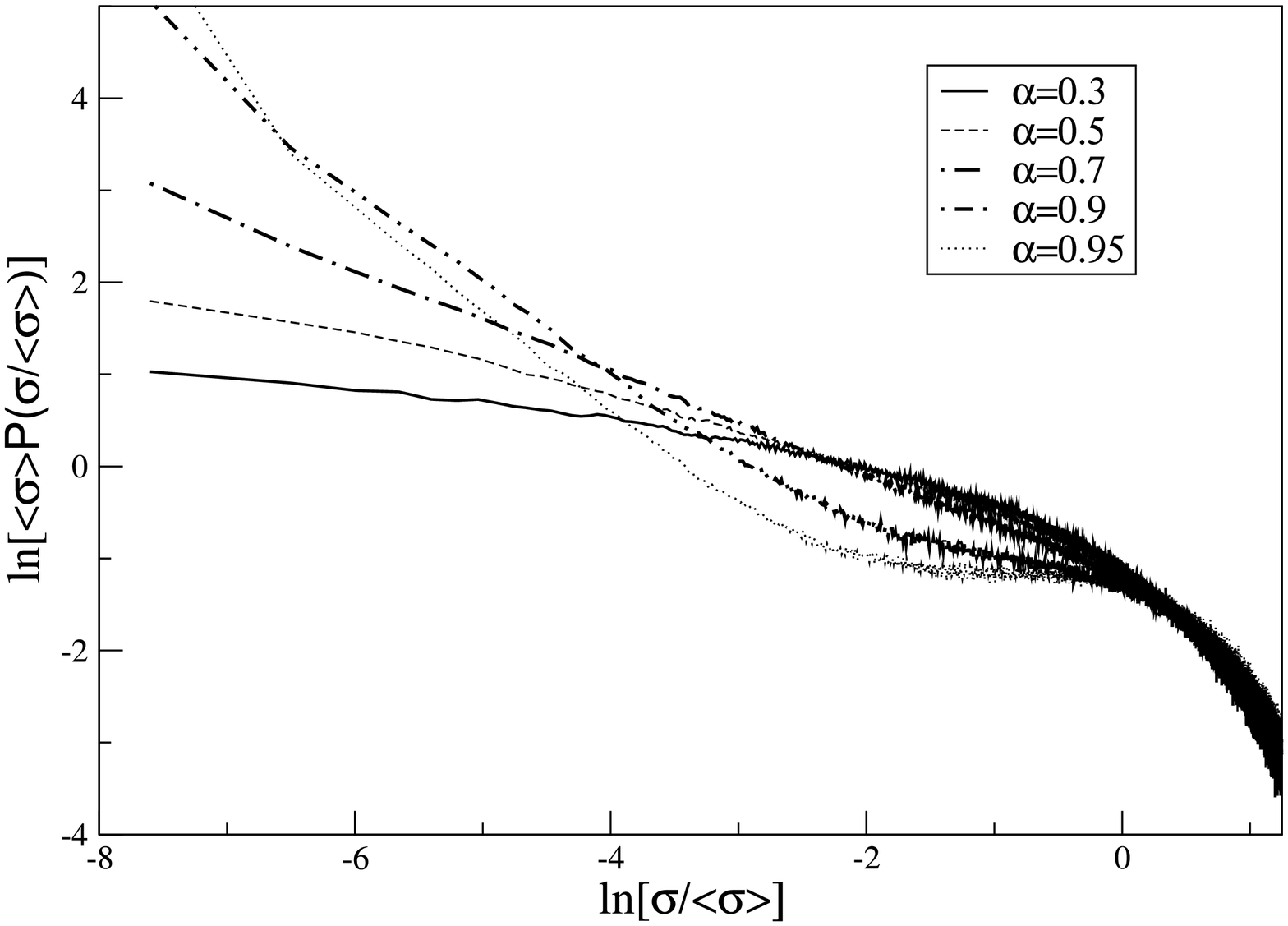}
\end{center}
\end{figure}
\vspace{1cm}
\newpage
\begin{figure}
\begin{center}
\includegraphics[width=.75\textwidth]{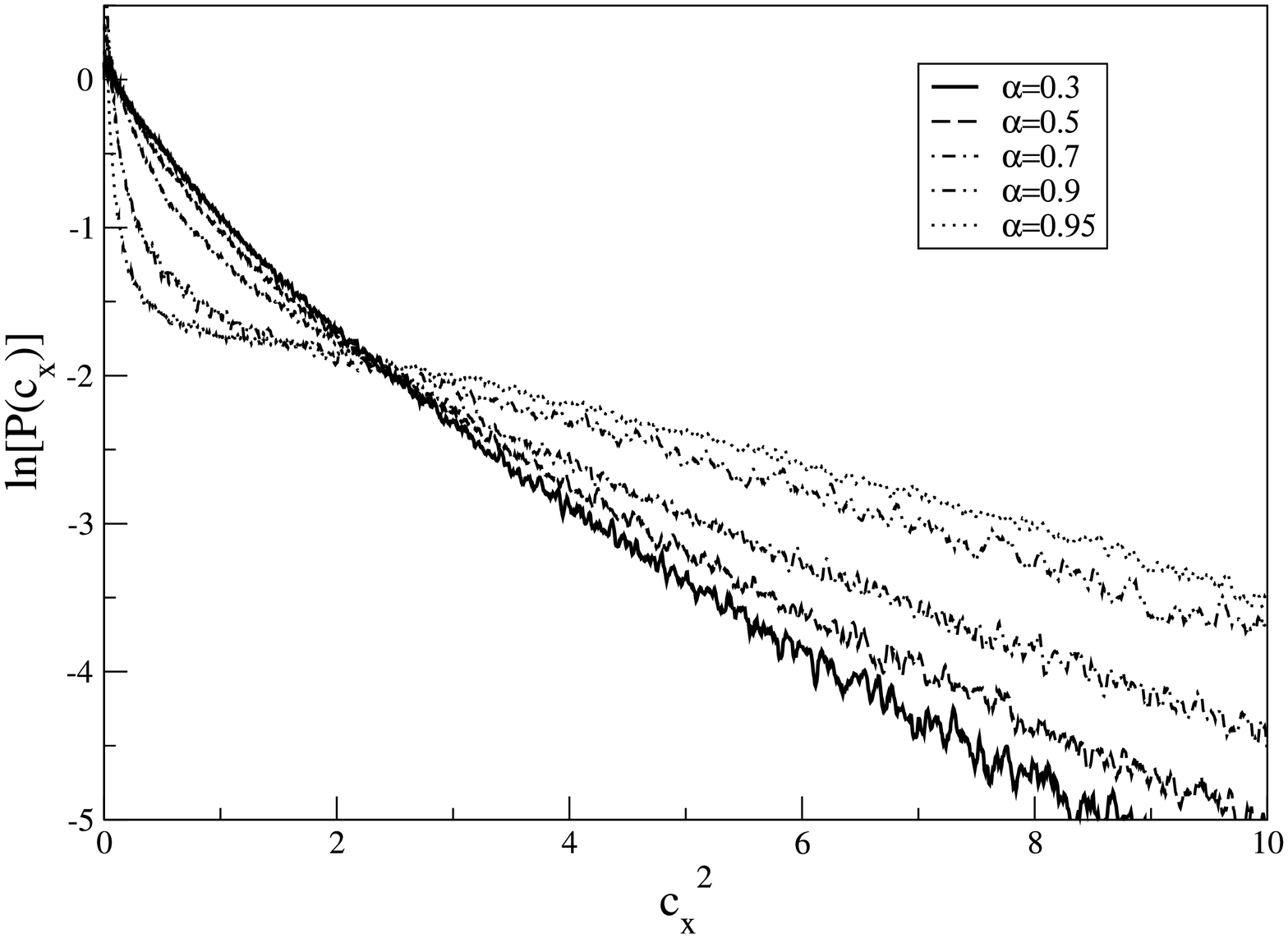}
\end{center}
\end{figure}
\vspace{1cm}
\newpage
\begin{figure}
\begin{center}
\includegraphics[width=.75\textwidth]{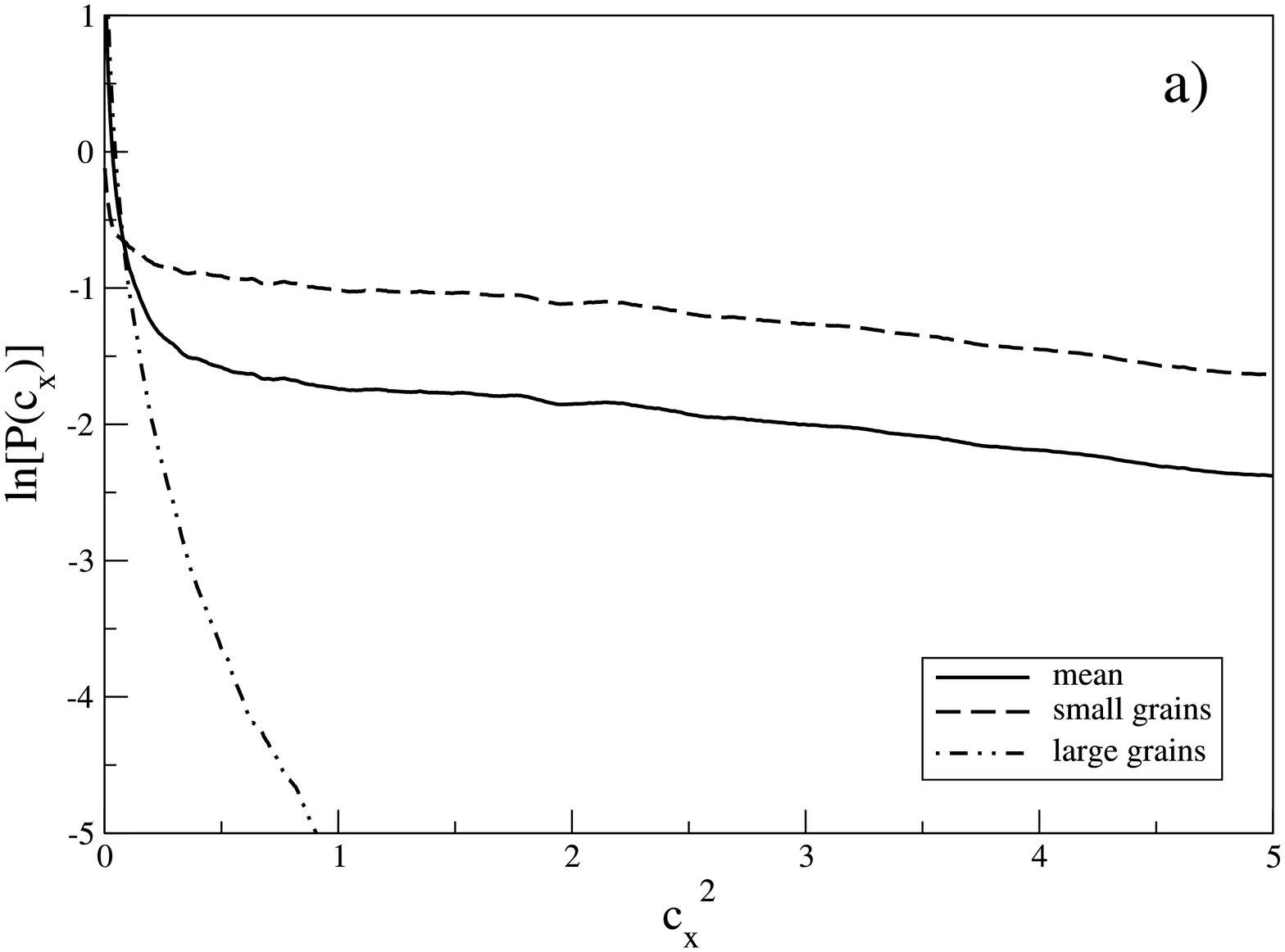}
\end{center}
\end{figure}
\begin{figure}
\begin{center}
\includegraphics[width=.75\textwidth]{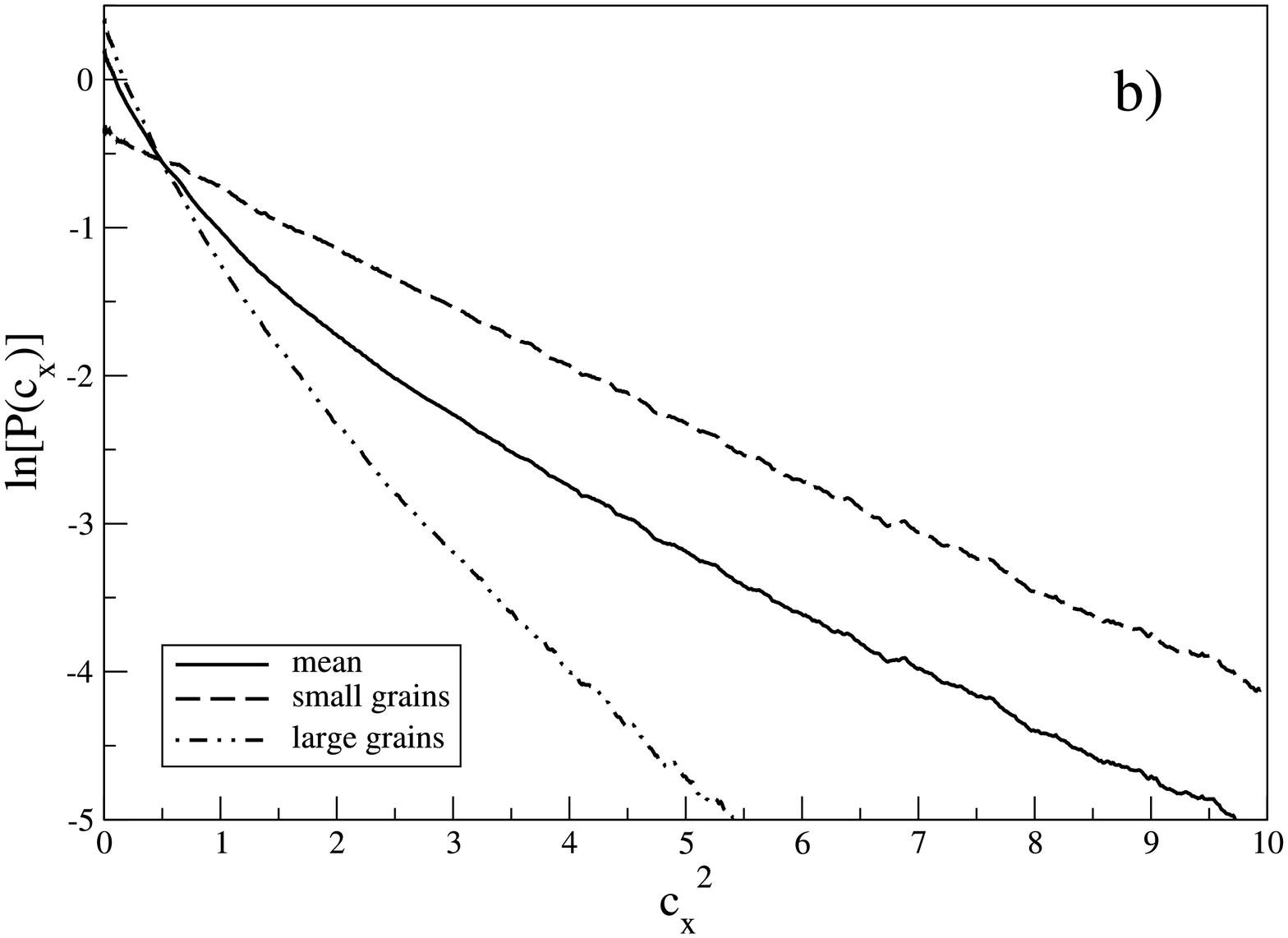}
\end{center}
\end{figure}
\vspace{1cm}
\end{document}